# Non-thermal fluence threshold for femtosecond pulsed x-ray radiation damage in perovskite complex oxide epitaxial heterostructures

Hyeon Jun Lee[1], Youngjun Ahn[1], Samuel D. Marks[1], Eric C. Landahl[2], Jun Young Lee[3], Tae Yeon Kim[3], Sanjith Unithrattil[3], Ji Young Jo[3], Sae Hwan Chun[4], Sunam Kim[4], Sang-Yeon Park[4], Intae Eom[4], Carolina Adamo[5], Darrell G. Schlom,[6,7] Haidan Wen[8], and Paul G. Evans[1,*]

[1] Department of Materials Science and Engineering, University of Wisconsin-Madison, Madison, Wisconsin 53706, USA

[2] Department of Physics, DePaul University, Chicago, Illinois 60614, USA

[3] School of Materials Science and Engineering, Gwangju Institute of Science and Technology, Gwangju 61005, South Korea

[4] Pohang Accelerator Laboratory, Pohang, Gyeongbuk 37673, South Korea

[5] Department of Applied Physics, Stanford University, Stanford, California 94305, USA

[6] Department of Materials Science and Engineering, Cornell University, Ithaca, New York 14853, USA

[7] Kavli Institute at Cornell for Nanoscale Science, Ithaca, New York 14853, USA

[8] X-ray Science Division, Argonne National Laboratory, Argonne, Illinois 60439, USA

Intense hard x-ray pulses from a free-electron laser induce irreversible structural damage in a perovskite oxide epitaxial heterostructure when pulse fluences exceed a threshold value. The intensity of x-ray diffraction from a 25-nm thick epitaxial BiFeO$_3$ layer on a SrTiO$_3$ substrate measured using a series of pulses decreases abruptly with a per-pulse fluence of $2.7 \times 10^6$ photons µm$^{-2}$ at 9.7 keV photon energy, but remains constant for $1.3 \times 10^6$ photons µm$^{-2}$ or less. The damage resulted in the destruction of the BiFeO$_3$ thin film within the focal spot area and the formation of a deep cavity penetrating into the STO substrate via the removal of tens of nanometers of material per pulse. The damage threshold occurs at a fluence that is insufficient to heat the absorption volume to the melting point. The morphology of the ablated sample is consistent with fracture rather than melting. Together these results indicate that the damage occurs via a non-thermal process consistent with ultrafast ionization of the absorption volume.

X-ray free electron lasers (XFELs) are powerful tools for the study of ultrafast phenomena and present new directions in probing the fundamental interaction between x-ray photons and condensed matter.[1-3] The x-ray pulses produced by XFELs have durations of tens of femtoseconds combined with extremely high brilliance and per-pulse energy and thus can have high x-ray fluence in µm-scale focal areas.[4,5] The intensities of focused XFEL beams reach a previously inaccessible regime of x-ray/matter interactions, in which the structural effects associated with a rapid adiabatic temperature rise and large photoinduced charge density have not yet been explored in detail. X-ray induced modifications of solid materials are important in understanding the x-ray dose limits that constrains XFEL experiments in condensed matter physics and materials science. More generally, understanding of x-ray/matter interactions using experiments at XFELs also has the potential to impact other fields involving intense pulses generated by a variety of x-ray sources, for example in inertial confinement fusion experiments.[6]

The absorption of intense x-ray pulses produces effects that can lead to permanent structural or chemical changes: adiabatic heating and melting, photoelectric absorption and ionization, and photochemical degradation. The extent of x-ray beam-induced damage varies as a function of the x-ray fluence per pulse and often occurs with a threshold fluence that depends on the cross sections for x-ray absorption.[7-9] For example, the damage thresholds evaluated by examining surface morphological changes in Si and Pt are 780 and 23 nJ µm$^{-2}$, respectively, a difference of more than an order of magnitude linked to the higher absorption constant of Pt.[7] In these cases, the sample is degraded by an effectively instantaneous adiabatic heating in which there is no heat transfer from the region of x-ray absorption to its surroundings because thermal diffusion is negligible during the femtosecond duration of XFEL pulses.[10] Damage due to adiabatic heating occurs when the per-pulse fluence exceeds the threshold value associated with the melting or vaporization.[11]

In addition, damage can occur due to mechanisms other than melting or vaporization.[12] For example, tungsten diffractive optics are damaged by pulsed x-ray radiation at 9 nJ μm$^{-2}$ due to mechanical fracture through repeated thermal stress, at far below fluence of 50 nJ μm$^{-2}$ predicted for the melting.[13] In the visible and near-visible photon energy regime, non-thermal damage of dielectric materials in femtosecond optical laser pulses occurs via avalanche ionization when there is insufficient time for absorbed energy to be coupled to the lattice.[14]

We focus here on the x-ray fluence regime relevant to experiments probing the dynamics of complex metal oxide thin films and nanostructures, in which experiments with tightly focused micron-scale x-ray beams yield diffracted x-ray count rates of 100 to 1000 photons per XFEL pulse. The absorption of hard x-rays occurs over depths of hundreds of nanometers to micrometers in transition-metal oxides, establishing a key experimental length scale. We report experiments in which XFEL pulses with x-ray fluence above the damage threshold degrade a BiFeO$_3$ (BFO) thin film and the SrTiO$_3$ (STO) substrate. The threshold is a factor of at least 70 lower than the predicted fluence for melting, indicating a non-thermal effect induces the x-ray damage of BFO and STO.

The XFEL experiment was performed at the X-ray Scattering & Spectroscopy end-station of the PAL-XFEL using the experimental geometry in Fig. 1(a).[15] The x-ray pulse repetition rate was 30 Hz, allowing the x-ray absorption volume to return to thermal conditions close to the initial state between pulses. A monochromatic x-ray beam with 9.7 keV photon energy was focused to a spot with a 10 μm full width at half maximum (FWHM) diameter using Be compound refractive lenses that were located 8.3 m upstream from sample. Diffracted x-rays were detected using a multi-port charge coupled device (MPCCD) detector operated in a regime in which the absolute number of diffracted photons could be determined for each x-ray pulse.[16] The incident x-ray fluence was varied by inserting 100 μm-thick Al attenuators into the incident beam, with a total attenuator thickness of up to 900 μm and transmission as low as $1.1 \times 10^{-3}$.

The sample consisted of an (001)-oriented BFO film with a thickness of 25 nm grown by reactive molecular-beam epitaxy on an STO substrate. X-ray pulses were incident at an angle $\theta$ of 17.5 to 18.5° with respect to the surface, near the Bragg condition for the BFO 002 reflection. The x-ray footprint on the sample surface along the incident beam direction was thus elongated by approximately a factor of 3.

The mean incident x-ray fluence was $F_I = <E_{FEL}>/A$, where $A$ is the FWHM area of the focused x-ray beam on the surface and $<E_{FEL}>$ is the mean number of x-ray photons per pulse. Two measurements gave similar and consistent values of the fluence. First, a calibrated beam position monitor gave $<E_{FEL}> = 2.2 \times 10^9$ photons pulse$^{-1}$. Second, the mean count rate at the peak of the BFO 002 reflection was 650 photons pulse$^{-1}$ with incident beam attenuators set for a transmission of $1.1 \times 10^{-3}$, giving $<E_{FEL}> = 3.1 \times 10^9$, based on the previously measured BFO 002 reflectivity. The unattenuated focused beam based on the diffraction measurement thus had $F_I = 1.2 \times 10^7$ photons μm$^{-2}$ pulse$^{-1}$.

The probability distribution of the total number of photons per XFEL pulse after monochromatization and focusing and before the attenuators, $E_{FEL}$, is shown in Fig. 1(b) for $3.6 \times 10^4$ pulses. The incident intensity of each pulse was determined by measuring the diffracted fluence of each pulse at the peak of the BFO 002 reflection and converting this to the corresponding incident fluence using the method described above. The observed distribution is accurately described by the self-amplified spontaneous emission pulse intensity distribution $p(E_{FEL})$:[17]

$$p(E_{FEL}) = \frac{M^M}{\Gamma(M)} \left(\frac{E_{FEL}}{<E_{FEL}>}\right)^{M-1} \frac{1}{<E_{FEL}>} \exp\left(\frac{-M\, E_{FEL}}{<E_{FEL}>}\right). \qquad (1)$$

Here $\Gamma(M)$ is the gamma function and $M$ is the number of longitudinal optical modes given by M

$= \langle(E_{FEL} - \langle E_{FEL}\rangle)^2\rangle / \langle E_{FEL}\rangle^2 + 1$. The dashed curve in Fig. 1(b) is plotted with $M = 4$. The use of the mean fluence $\langle E_{FEL}\rangle$ to parameterize the experimental results is an important detail that may lead to uncertainty in the precise value of the damage threshold because the series of pulses during the experiment includes pulses with fluence far higher than $\langle E_{FEL}\rangle$.

Radiation damage to BFO/STO film was probed using diffraction pattern of the BFO layer. The initial thin film diffraction pattern near the BFO 002 reflection acquired with a total of 4600 XFEL pulses with $F_I = 1.8 \times 10^4$ photons μm$^{-2}$ pulse$^{-1}$ is shown in Fig. 1(c). Thickness fringes appear in Fig. 1(c) with a spacing of 0.025 Å$^{-1}$, matching the BFO film thickness of 25 nm.

Fluences below the damage threshold did not produce systematic changes in the thin film diffraction pattern. Diffraction profiles near the BFO reflection acquired with $F_I = 1.3 \times 10^4$, $2.7 \times 10^5$, and $1.3 \times 10^6$ photons μm$^{-2}$ pulse$^{-1}$ are shown in Fig. 2(a). Each $Q_z$ point of each diffraction pattern was acquired using 10 pulses and normalized using a weighted average based on the fluence of each pulse. The total number of pulses for each pattern was between $1 \times 10^4$ and $2 \times 10^4$, including pulses arriving during the motion of the diffractometer. The diffraction patterns over the entire range of intensities in Fig. 2(a) are accurately fit by the same parameters used for the initial pattern in Fig. 1(c).

The BFO diffraction pattern arises only from the epitaxial BFO layer and thus provides a precise signature of surface damage. At the lowest fluence, $F_I = 1.3 \times 10^4$ photons μm$^{-2}$ pulse$^{-1}$, the diffraction pattern was acquired at the peak of the BFO 002 reflection. The long-term evolution of the diffracted intensity with more intense beams was measured using thickness fringes at $Q_z = 3.057$ Å$^{-1}$ for $F_I = 2.7 \times 10^5$ photons μm$^{-2}$ pulse$^{-1}$ and at $Q_z = 3.032$ Å$^{-1}$ for $F_I = 1.3 \times 10^6$ photons μm$^{-2}$ pulse$^{-1}$ in order to keep the number of diffracted x-ray photons from exceeding the maximum per-pixel detected fluence at the detector. The diffracted intensities tracked for between $2 \times 10^4$ and 1

× $10^5$ pulses are shown in Fig. 2(b). The range of $Q_z$ over which diffraction data was collected at each fluence was selected in order to keep the number of diffracted x-ray photons from exceeding the maximum per-pixel detected fluence at the detector. The diffracted intensity is independent of the number of pulses for mean fluences up to $F_I = 1.3 \times 10^6$ photons μm$^{-2}$ pulse$^{-1}$, indicating that the BFO layer was not destroyed by x-ray pulses at or below this mean fluence.

There was a rapid degradation of the BFO layer at high fluence, for $F_I = 2.7 \times 10^6$ photons μm$^{-2}$ pulse$^{-1}$ and above, resulting in a significant decrease in the diffracted intensity. The diffracted intensity measured at the peak of 002 reflection for $F_I$=2.7 × $10^6$ and 5.9 × $10^6$ photons μm$^{-2}$ pulse$^{-1}$, shown in Fig. 2(b), is lower than for the undamaged BFO layer and results only from the spatial overlap of the low-intensity tail of the Gaussian focused beam with undamaged regions of the BFO film. This diffraction by the tails of the focused beam produces a measurable diffracted intensity even when the BFO layer in the central focal area has been completely destroyed. The diffraction intensities were not recorded for exposure to individual pulses at large $F_I$, but we hypothesize that the BFO layer was immediately degraded after exposure to a small number of pulses.

The morphological changes resulting from x-ray pulses with fluence above the damage threshold were studied using scanning electron microscopy (SEM) and optical microscopy. An SEM image of film surface after 3.6 × $10^3$ pulses with $F_I$ = 5.9 × $10^6$ photons μm$^{-2}$ pulse$^{-1}$ is shown in Fig. 3(a). The x-ray pulses destroy the area of the BFO film and the underlying STO substrate in the region of the central x-ray focus, yielding a cylindrical hole with 12 μm diameter that matches the spot size of the focused x-ray pulse. The hole penetrates the BFO layer and the substrate to a depth of several hundred micrometers, with a direction matching the x-ray incident angle.

The perimeter of the hole in Fig. 3(a) exhibits a brittle fracture pattern consisting of curved surface features. A similar pattern is generated at fracture surfaces due to concentrated stress in ceramics.[18] The curved pattern repeats at different depths, indicating that the stress leading to the damage was applied at different depth during a series of pulses. The fracture surface observed in Fig. 3(a) is completely different from the melting observed in samples damaged by heating effects.[19] The optical image focused at the surface, in the upper panel of Fig. 3(b), reveals a structure similar to the SEM image in Fig. 3(a). Focusing the optical microscope beneath the surface of the transparent substrate, lower panel of Fig. 3(b), reveals a cavity starting from the surface and continuing inside the STO substrate. The total length of the cavity produced via x-ray beam damage after $3.6 \times 10^3$ XFEL pulses with $5.9 \times 10^6$ photons µm$^{-2}$ pulse$^{-1}$ was 740 µm, corresponding to an average thickness removed by each x-ray pulse at this fluence of 200 nm. The series of $1.8 \times 10^4$ pulses with a lower fluence $2.7 \times 10^6$ photons µm$^{-2}$ pulse$^{-1}$ removed 820 µm, equivalent to 45 nm pulse$^{-1}$. The depth of removed material per pulse is shown as a function of pulse fluence in Fig. 4.

The adiabatic heating damage mechanism can be quantitatively considered by predicting the temperature increase $\Delta T$ per pulse within the x-ray absorption volume. The characteristic time for cooling, estimated using the size of the beam and the thermal diffusivity of STO, is on the order of µs.[11] This cooling time is much shorter than the interval between x-ray pulses, indicating that there is a negligible cumulative increase in the temperature during the total time of the series of x-ray pulses. Under adiabatic heating conditions, $\Delta T = F_I E_P \mu/\rho C_p$, where $E_P$ is the incident x-ray photon energy, $\rho$ is the mass density, $\mu$ is the x-ray absorption coefficient, and $C_p$ is the specific heat. The absorption coefficients BFO and STO, at 9.7 keV are 1167 cm$^{-1}$ and 222 cm$^{-1}$ and the of $C_p$ for BFO and STO are 120 J mol$^{-1}$ K$^{-1}$ and 100 J mol$^{-1}$ K$^{-1}$.[20,21] The values of $\rho$ for BFO and STO are 8.41g cm$^{-3}$ and 5.10 g cm$^{-3}$, respectively. The temperature increases for BFO layer and STO are thus 159 K and 33 K, respectively, for $F_I = 2.7 \times 10^6$ photons µm$^{-2}$ pulse$^{-1}$, at which damage is unambiguously observed. The temperatures reached by adiabatic heating are thus far lower than melting temperatures of BFO and STO, 1235 K and 2350 K, respectively.[22,23] The additional contribution of the latent heat required to melt or otherwise transform the sample would result in an even higher required fluence for thermal damage. We conclude that FEL pulses with the mean fluence do not provide sufficient energy for irreversible transformation in the sample via heating to the melting point. It is in principle possible that SASE pulses with fluence many times higher than the average, would result in heating above the melting point or other comparatively high benchmark temperatures for beam damage. However, we unambiguously and repeatedly observe damage with far fewer pulses than would be statistically required to yield such a high-intensity pulse and we thus conclude that the damage mechanism is not due to heating.

The pattern of the damaged surface and depth removed per pulse are consistent with rapid localized charging and associated mechanical degradation, often termed a Coulomb explosion.[14] The process of degradation begins with the excitation of a high charge density that cannot readily be recombined because insulating BFO and STO do not permit charge transport to the ionized atoms.[24,25] Photoelectrons escape from a depth on the order of the range determined by their kinetic energy and excite secondary electrons with a wide range of energies. These processes also occur in photoelectron-based materials analysis techniques, including x-ray photoelectron spectroscopy. The kinetic energy of excited electrons from L- and M-shells is on the order of 5 keV, for which the range given by the Kanaya-Okayama approximation is approximately 1 µm in STO.[26] This range corresponds to the order of magnitude for the depth from which electrons can escape the STO, but should be treated as an approximation.[26] The electrons escaping from the material leave positively charged ions and generate a near-surface electric field. At higher incident x-ray fluence, the higher number of escaping electrons leads to an increased magnitude of generated electric field.[27] This damage process is conceptually similar to laser ablation, e.g. as employed in pulsed laser deposition, in which optical pulses with short duration lead to impact ionization, dielectric breakdown, and large-scale transport of ablated material.[14] We hypothesize that the experimentally observed sharp damage threshold results from the fluence at which the x-ray induced electric field near the surface exceeds the dielectric breakdown field or a similar critical value.

This quantification of the damage threshold and mechanism can have a significant impact in the design of pulsed x-ray studies of complex oxide materials. Time-resolved studies of structural transients in epitaxial complex oxides take advantage of the extremely intense and short duration of the pulses of x-ray radiation produced by XFELs, but require multiple pulses and must avoid sample damage.[28,29] More generally, the potential use of focused x-ray beams with fluences below the damage threshold can be employed for time-resolved x-ray microscopy and coherent diffraction imaging experiments of the dynamics of heterogeneous materials under external stimuli can permit the study of isolated features or nanoscale devices.[30,31]

The authors gratefully acknowledge support from the U.S. DOE, Basic Energy Sciences, Materials Sciences and Engineering, under contract no. DE-FG02-04ER46147. The authors gratefully acknowledge use of facilities and instrumentation at the UW-Madison Wisconsin Centers for Nanoscale Technology (wcnt.wisc.edu) partially supported by the NSF through the University of Wisconsin Materials Research Science and Engineering Center (DMR-1720415). The experiment was performed at the XSS beamline of PAL-XFEL (proposal no. 2018-1st-XSS-003) funded by the Ministry of Science and ICT of Korea. The work at Cornell University was supported by the National Science Foundation (Nanosystems Engineering Research Center for Translational Applications of Nanoscale Multiferroic Systems) under grant number EEC-1160504 (C.A. and D.G.S.). This work was performed in part at the Cornell Nanoscale Facility, a member of the National Nanotechnology Coordinated Infrastructure (NNCI), which is supported by the National Science Foundation (Grant ECCS-1542081).

**Figure 1** (a) Schematic of diffraction experiment, including Al attenuators, and normalization photodiode. (b) Histogram pulse intensities for $3.6 \times 10^4$ FEL x-ray pulses. The dashed curve is the SASE distribution predicted by eq. (1) with $M = 4$. (c) Initial diffraction pattern of the BFO 002 reflection with $1.8 \times 10^4$ photons $\mu m^{-2}$ pulse$^{-1}$. The dashed line shows the predicted diffraction pattern for a BFO thin film using a kinematical x-ray diffraction calculation.

**Figure 2** (a) Diffraction patterns of the 002 reflection of the BFO film acquired with a range of x-ray fluences. Intensities were normalized to the peak intensity of 002 reflection measured with $1.3 \times 10^4$ photons $\mu m^{-2}$ pulse$^{-1}$. (b) Normalized intensities of the features of the BFO diffraction pattern as a function of the total number of pulses: (squares) 002 reflection with $1.3 \times 10^4$ photons $\mu m^{-2}$ pulse$^{-1}$, (red circles) at $Q_z=3.057$ Å$^{-1}$ with $2.7 \times 10^5$ photons $\mu m^{-2}$ pulse$^{-1}$, and (green triangle) at $Q_z=3.057$ Å$^{-1}$ with $1.3 \times 10^6$ photons $\mu m^{-2}$ pulse$^{-1}$. The residual intensities represented in shaded region after damage with (blue triangles) $2.7 \times 10^6$ and (purple diamonds) $5.9 \times 10^6$ photons $\mu m^{-2}$ pulse$^{-1}$ were measured from 002 reflection.

**Figure 3** (a) SEM image of damaged region after 3600 pulses at $F_I = 5.9 \times 10^6$ photons $\mu m^{-2}$ pulse$^{-1}$. (c) Optical microscopy images of damaged area acquired with the optical focal plane at the surface (upper panel) and beneath the surface of the transparent sample (lower panel). Arrows indicate the location of the damaged area in the focal plane at the surface and within the STO substrate, respectively.

**Figure 4** Fluence dependence of the depth of the BFO/STO heterostructure removed per XFEL pulse.